\documentclass{proceedingsM}
\usepackage{times,amsmath,amssymb}
\usepackage{graphicx,color}
\usepackage{cases}

\newcommand{\ds}{\displaystyle}

\title{Numerical investigation of acoustic solitons}

\author{
Bruno Lombard\footnotemark[1], 
Jean-Fran\c{c}ois Mercier\footnotemark[2],
Olivier Richoux\footnotemark[3]
}
\address{{\footnotemark[1]} LMA, CNRS, UPR 7051, Aix-Marseille Universit\'e, Centrale Marseille, 13402 Marseille, France}
\address{{\footnotemark[2]} POEMS, CNRS/ENSTA/INRIA, UMR 7231, ENSTA ParisTech, 91762 Palaiseau, France}
\address{{\footnotemark[3]} LAUM, UMR 6613 CNRS, Universit\'e du Maine, 72085 Le Mans, France}
\email{lombard@lma.cnrs-mrs.fr, jean-francois.mercier@ensta.fr, olivier.richoux@univ-lemans.fr}

\abstract{
Acoustic solitons can be obtained by considering the propagation of large amplitude sound waves across a set of Helmholtz resonators. The model proposed by Sugimoto and his coauthors has been validated experimentally in previous works. Here we examine some of its theoretical properties: low-frequency regime, balance of energy, stability. We propose also numerical experiments illustrating typical features of solitary waves.
}


\begin{document}

\maketitle

%---------------------------------------------------------------------------------

\section{Introduction}\label{SecIntro}

Solitons are nonlinear waves with large amplitude and constant profile, resulting from the competition between nonlinearity and dispersion. They occur in many physical area, such as fluid mechanics (Korteweg-de Vries equations), electromagnetism and optics (Klein-Gordon equations) \cite{Engelbrecht88}. In acoustics, the intrinsic dispersion is too low compared to the nonlinearity to produce solitons. Thus additional geometric dispersion must be considered to observe acoustic solitons. It was the basis of a series of works of Sugimoto and coauthors \cite{Sugimoto92,Sugimoto04}, where the propagation of shock waves was investigated in a tube connected to an array of Helmholtz resonators. A mathematical model was proposed, as well as a theoretical analysis and a comparison with experimental data.

Sugimoto's work was extended in two means. In \cite{Lombard14}, a time-domain numerical model was proposed to incorporate efficiently the fractional derivatives modeling linear viscothermic losses. In \cite{Richoux15}, comparisons with experimental results were proposed. It was shown that nonlinear attenuation in the resonators had also to be incorporated for describing accurately the experiments. 

The goal of the present contribution is to analyse further the full Sugimoto's model with fractional derivatives and nonlinear attenuation, recalled in section \ref{Sugimoto} In the low-frequency regime, corresponding to the experimental conditions, the evolution equations tend to a Korteweg-de Vries equation with an additional nonlinear term. Therefore one expects solitons nonlinearly attenuated. Both for mathematical and numerical purposes, the fractional model is transformed by means of a diffusive representation (section \ref{SecHyp}). Doing so allows to analyse the energy balance and the stability of the model (sections \ref{SecNrj} and \ref{SecStability}). Lastly, two sets of numerical experiments are proposed in section \ref{SecNum}, showing that the waves have the typical features of solitary waves.

%---------------------------------------------------------------------------------
%---------------------------------------------------------------------------------

\section{Fractional model}\label{SecEDP}

\subsection{Sugimoto's equations}\label{Sugimoto}

The configuration is depicted in figure 2 of \cite{Lombard14}. The wavelengths are much larger than the distance between two resonators, so that the latter are described by a continuous distribution. One-dimensional propagation is assumed. The variables are the velocity of gas $u$ and the excess pressure in the resonators $p$. Considering the right-going wave, one writes \cite{Sugimoto92}
\begin{subnumcases}{\label{EDP}}
\displaystyle
\frac{\textstyle \partial u}{\textstyle \partial t}+\frac{\textstyle \partial}{\textstyle \partial x}\left(a \textstyle u +b\displaystyle \frac{\textstyle u^2}{\textstyle 2}\right)= c\frac{\textstyle \partial^{-1/2}}{\textstyle \partial t^{-1/2}}\frac{\textstyle \partial u}{\textstyle \partial x}+d\frac{\textstyle \partial^2 u}{\textstyle \partial x^2}-e\frac{\textstyle \partial p}{\textstyle \partial t},\label{EDP1}\\
%[8pt]
\displaystyle
\frac{\textstyle \partial^2 p}{\textstyle \partial t^2}+f\frac{\textstyle \partial^{3/2} p}{\textstyle \partial t^{3/2}}+gp-m\frac{\textstyle \partial^2 (p)^2}{\textstyle \partial t^2}+n\left|\frac{\textstyle \partial p}{\textstyle \partial t}\right|\,\frac{\textstyle \partial p}{\textstyle \partial t}= hu\label{EDP2}.
\end{subnumcases}
The PDE (\ref{EDP1}) describes the nonlinear wave propagation (coefficients $a$ and $b$) in the tube. The losses in the tube are introduced by $c$ (viscothermic losses at the wall) and by $d$ (volume attenuation). The ODE (\ref{EDP2}) describes the oscillations in the resonators. In the latter, the losses are introduced by $f$ (viscothermic losses), by $m$ and by $n$ (nonlinear attenuation due to turbulence). Coupling between (\ref{EDP1}) and (\ref{EDP2}) is ensured by $e$ and $h$. See \cite{Richoux15} for the expression of all these coefficients. A fractional integral of order 1/2 ($c$) and a fractional derivative of order 3/2 ($f$) are introduced. These non-local operators are tackled with in section \ref{SecHyp}

%---------------------------------------------------------------------------------

\subsection{Low-frequency approximation}\label{SecKdv}

Under the hypothesis of weak nonlinearity, $\partial u/\partial x$ in (\ref{EDP1}) is replaced by $-(1/a)\,\partial u/\partial t$ in the terms with coefficients $b$, $c$ and $d$. The resulting system is written in the $(T,\,X)$ coordinates, where $T$ is a non-dimensional retarded time, $X$ is a non-dimensional slow space variable: 
\begin{equation}
T=\omega\left(t-\frac{\textstyle x}{\textstyle a}\right),\qquad X=\varepsilon\,\omega\,\frac{\textstyle x}{\textstyle a},\qquad \varepsilon=\frac{\textstyle \gamma+1}{\textstyle 2}\,\frac{\textstyle u_{\max}}{\textstyle a},
\label{ThetaX}
\end{equation}
where $u_{\max}$ is the magnitude of the gas velocity at the initial time, $\omega$ is a characteristic wave frequency, and $\gamma$ is the ratio of specific heats at constant pressure and volume. Introducing the reduced variables $U$ and $P$
\begin{equation}
U=\frac{\textstyle 1}{\textstyle \varepsilon}\,\frac{\textstyle \gamma+1}{\textstyle 2}\,\frac{\textstyle u}{\textstyle a}={\cal O}(1),\quad
P=\frac{\textstyle 1}{\textstyle \varepsilon}\,\frac{\textstyle \gamma+1}{\textstyle 2\,\gamma}\,\frac{\textstyle p}{\textstyle p_0}={\cal O}(1),
\label{UP}
\end{equation}
where $p_0$ is the pressure at equilibrium, one obtains the system
\begin{subnumcases}{\label{EDPscale}}
\displaystyle
\frac{\textstyle \partial U}{\textstyle \partial X}-U\,\frac{\textstyle \partial U}{\textstyle \partial T}=-\delta_R\,\frac{\textstyle \partial^{1/2} U}{\textstyle \partial T^{1/2}}+\beta\,\frac{\textstyle \partial^2 U}{\textstyle \partial T^2}-K\,\frac{\textstyle \partial P}{\textstyle \partial T},\label{EDPscale1}\\
%[8pt]
\displaystyle
\frac{\textstyle \partial^2 P}{\textstyle \partial T^2}+\delta_r\,\frac{\textstyle \partial^{3/2} P}{\textstyle \partial T^{3/2}}+\Omega\,P-M\frac{\textstyle \partial^2 P^2}{\textstyle \partial T^2}+N\,\left|\frac{\textstyle \partial P}{\textstyle \partial T}\right|\,\frac{\textstyle \partial P}{\textstyle \partial T}=\Omega\,U.\label{EDPscale2}
\end{subnumcases}
This system generalizes the equations (2-5) and (2-6) of \cite{Sugimoto04} to the case of nonlinear losses (terms with $M$ and $N$). As shown in \cite{Sugimoto92}, $\beta$ is negligible compared to $\delta_r$ and $\delta_R$. We consider waves with characteristic frequencies much smaller than the natural frequency of the resonators $\omega_e$, so that $\Omega=(\omega_e/\omega)\gg 1$. In this case, the dispersion analysis performed in \cite{Lombard14} indicates that the viscothermal losses are small. Moreover, the volume of the resonators is large compared to that of the necks, so that $M\ll N$ is neglected. Consequently, the low-frequency regime $\Omega\gg 1$ yields the simplified system
\begin{subnumcases}{\label{EDPscaleOm}}
\displaystyle
\frac{\textstyle \partial U}{\textstyle \partial X}-U\,\frac{\textstyle \partial U}{\textstyle \partial T}=-K\,\frac{\textstyle \partial P}{\textstyle \partial T},\label{EDPscaleOm1}\\
%[8pt]
\displaystyle
\frac{\textstyle \partial^2 P}{\textstyle \partial T^2}+\Omega\,P+N\,\left|\frac{\textstyle \partial P}{\textstyle \partial T}\right|\,\frac{\textstyle \partial P}{\textstyle \partial T}=\Omega\,U.\label{EDPscaleOm2}
\end{subnumcases}
From (\ref{EDPscaleOm2}), one obtains
\begin{equation}
\begin{array}{lll}
P &=& \displaystyle U-\frac{\textstyle 1}{\textstyle \Omega}\frac{\textstyle \partial^2 U}{\textstyle \partial T^2}-\frac{\textstyle N}{\textstyle \Omega}\,\left|\frac{\textstyle \partial U}{\textstyle \partial T}\right|\,\frac{\textstyle \partial U}{\textstyle \partial T}+{\cal O}\left(\frac{\textstyle 1}{\textstyle \Omega^2}\right).
\end{array}
\label{Omega-1}
\end{equation}
Injecting (\ref{Omega-1}) in (\ref{EDPscaleOm1}) gives:
\[
\frac{\textstyle \partial U}{\textstyle \partial X}+K\,\frac{\textstyle \partial U}{\textstyle \partial T}-U\,\frac{\textstyle \partial U}{\textstyle \partial T}=\frac{\textstyle K}{\textstyle \Omega}\frac{\textstyle \partial^3 U}{\textstyle \partial T^3}+\frac{2 \textstyle K\,N}{\textstyle \Omega}\,\left|\frac{\textstyle \partial U}{\textstyle \partial T}\right|\,\frac{\textstyle \partial^2 U}{\textstyle \partial T^2}+{\cal O}\left( \frac{1}{\Omega^2} \right).
\]
Neglecting the second-order terms in $1/\Omega$ and introducing the new unknown $V=U-K$ leads to the PDE
\begin{equation}
\frac{\textstyle \partial V}{\textstyle \partial X}-V\,\frac{\textstyle \partial V}{\textstyle \partial T}=\frac{\textstyle K}{\textstyle \Omega}\frac{\textstyle \partial^3 V}{\textstyle \partial T^3}+\frac{2 \textstyle K\,N}{\textstyle \Omega}\,\left|\frac{\textstyle \partial V}{\textstyle \partial T}\right|\,\frac{\textstyle \partial^2 V}{\textstyle \partial T^2}.
\label{Korteweg}
\end{equation}
When nonlinear attenuation in the resonators is neglected ($N=0$), equation (\ref{Korteweg}) recovers the Korteweg-de Vries equation (2-35) of \cite{Sugimoto92}, which allows the propagation of solitons. Solitons are also expected to exist for small $N$ values, but with a decrease of amplitude. 

%---------------------------------------------------------------------------------
%---------------------------------------------------------------------------------

\section{Diffusive model}\label{SecDiff}

\subsection{Evolution equations}\label{SecHyp}

A diffusive approximation of the non-local in time fractional operators in (\ref{EDP}) is followed here \cite{Matignon08}. The half-order integral of a function $w(t)$ can be written 
\begin{equation}
\frac{\textstyle \partial^{-1/2}}{\textstyle \partial t^{-1/2}}w(t)=\int_0^{+\infty}\phi(t,\theta)\,d\theta\simeq
\sum_{\ell=1}^{N}\mu_{\ell}\,\phi_{\ell}(t),
\label{I12}
\end{equation}
where the diffusive variable $\phi$ satisfies the local-in-time ordinary differential equation
\begin{equation}
\frac{\partial \phi}{\partial t}=-\theta^2\,\phi+\frac{\textstyle 2}{\textstyle \pi}\,w.
\label{ODEI12}
\end{equation}
In (\ref{I12}), $\phi(t,\theta_\ell)=\phi_\ell(t)$; $\mu_{\ell}$ and $\theta_\ell$ are the weights and nodes of the quadrature formula. Their computation is detailed in \cite{Richoux15}. A similar derivation is applied to the 3/2 derivative in (\ref{EDP}), involving the diffusive variable $\xi$. Injecting these diffusive approximations in (\ref{EDP}) yields the following system of evolution equations
\begin{equation}
\left\{
\begin{array}{l}
\displaystyle
\frac{\textstyle \partial u}{\textstyle \partial t}+\frac{\textstyle \partial}{\textstyle \partial x}\left(a \textstyle u +b\displaystyle \frac{\textstyle (u)^2}{\textstyle 2}\right)=c\sum_{\ell=1}^{N}\mu_{\ell}\,\phi_{\ell}+d \frac{\textstyle \partial^2 u}{\textstyle\partial x^2}- e q,\\
%[10pt]
\displaystyle
\frac{\textstyle \partial p}{\textstyle \partial t}=q,\\
%[8pt]
\displaystyle
\frac{\textstyle \partial q}{\textstyle \partial t}=\frac{\textstyle 1}{\textstyle 1-2 m p}\left(h u-g p-f \sum_{\ell=1}^{N}\mu_{\ell}\left(-\theta_{\ell}^2\,\xi_{\ell}+\frac{\textstyle 2}{\textstyle \pi}\,q\right)+2 m (q)^2-n |q|\,q\right),\\
%[14pt]
\displaystyle
\frac{\textstyle \partial \phi_{\ell}}{\textstyle \partial t}-\frac{\textstyle 2}{\textstyle \pi}\,\frac{\textstyle \partial u}{\textstyle \partial x}=-\theta_{\ell}^2\,\phi_{\ell},\hspace{1cm} \ell=1\cdots N,\\
[12pt]
\displaystyle
\frac{\textstyle \partial \xi_{\ell}}{\textstyle \partial t}=-\theta_{\ell}^2\,\xi_{\ell}+\frac{\textstyle 2}{\textstyle \pi}\,q,\hspace{1.3cm} \ell=1\cdots N,
\end{array}
\right.
\label{SystComplet}
\end{equation}
The $(3+2\,N)$ unknowns are gathered in the vector
\begin{equation}
{\bf U}=\left(u,\,p,\,q,\phi_1,\cdots,\,\phi_{N},\,\xi_1,\cdots,\,\xi_{N}\right)^T.
\label{VecU}
\end{equation}
Then the nonlinear system (\ref{SystComplet}) can be written in the form
\begin{equation}
\frac{\textstyle \partial}{\textstyle \partial t}{\bf U}+\frac{\textstyle \partial}{\textstyle \partial x}{\bf F}({\bf U})={\bf S}({\bf U})+{\bf G}\,\frac{\textstyle \partial^2}{\textstyle \partial x^2}{\bf U}.
\label{SystHyper}
\end{equation}

%---------------------------------------------------------------------------------

\subsection{Energy balance}\label{SecNrj}

Based on the system (\ref{SystComplet}), we define
\begin{equation}
{\cal E}=\frac{\textstyle 1}{\textstyle 2}\int_\mathbb{R}\left(u^2+\frac{\textstyle eg}{\textstyle h}p^2+\frac{\textstyle e}{\textstyle h}q^2+\frac{\textstyle \pi}{\textstyle 2}\frac{\textstyle ef}{\textstyle h}\sum_{\ell=1}^{N}\mu_{\ell}\,\theta_{\ell}^2\,\xi_{\ell}^2\right)\,dx.
\label{NRJ}
\end{equation}
Assuming smooth solutions (no shock) and $c=0$, then one obtains
\begin{equation}
\begin{array}{l}
\ds
\frac{\textstyle d{\cal E}}{\textstyle dt}=-\int_\mathbb{R}d\left(\frac{\textstyle \partial u}{\textstyle \partial x}\right)^2dx-\frac{\textstyle \pi}{\textstyle 2}\int_\mathbb{R}\frac{\textstyle ef}{\textstyle h}\sum_{\ell=1}^{N}\mu_{\ell}\left(\frac{\textstyle \partial \xi_{\ell}}{\textstyle \partial t}\right)^2dx-\int_\mathbb{R}\frac{\textstyle en}{\textstyle h}|q|q^2 dx+\int_\mathbb{R}\frac{\textstyle e m}{\textstyle h}\frac{\textstyle \partial^2 p^2}{\textstyle \partial t^2}qdx.
\end{array}
\label{dEdt}
\end{equation}
It follows two remarks. First, if the weights of the diffusive approximation are positive $\mu_\ell>0$, then ${\cal E}$ is a quadratic definite positive form, which thus defines an energy. In practice, we determine these weights by an optimization procedure with constraint of positivity \cite{Richoux15}. Second, if the coefficient of nonlinear attenuation satisfies $m=0$, then $d{\cal E}/dt<0$: the energy decreases, and the model is well-posed. In practice, this hypothesis is reasonable, since $m\ll n$. 

Let us finally examine the assumed hypotheses. With shocks, the wave motion is irreversible and additional terms of dissipation must be accounted for in (\ref{dEdt}). On the other hand, the hypothesis $c=0$ is not physical but required for technical purpose: up to now, we have not found an energy if $c \neq 0$.

%---------------------------------------------------------------------------------

\subsection{Stability analysis}\label{SecStability}

The system (\ref{SystHyper}) is solved by a splitting technique \cite{Richoux15}: one successively solves the PDE
\begin{equation}
\frac{\textstyle \partial}{\textstyle \partial t}{\bf U}+\frac{\textstyle \partial}{\textstyle \partial x}{\bf F}({\bf U})={\bf G}\,\frac{\textstyle \partial^2}{\textstyle \partial x^2}{\bf U}
\label{SplitA}
\end{equation}
and the ODE
\begin{equation}
\frac{\textstyle \partial}{\textstyle \partial t}{\bf U}={\bf S}({\bf U})
\label{SplitB}
\end{equation}
with adequate time steps. Here we examine the stability of both stages. First, $\frac{\partial {\bf F}}{\partial {\bf U}}$ has real eigenvalues $\{a+b\,u, \,0^{2\,N+2}\}$ and is diagonalizable. Consequently, (\ref{SplitA}) is hyperbolic when ${\bf G}={\bf 0}$. In practice, ${\bf G}$ introduces a parabolic regularization, and the problem remains well-posed.

We have no general result about the stability of (\ref{SplitB}). But some partial results have been obtained, depending on the dissipation mechanisms considered:
\begin{itemize}
\item[(i)] nonlinear attenuation ($m\neq 0$ or $n \neq 0$), no fractional losses ($c=f=0$). Then the eigenvalues of ${\bf T}=\frac{\partial {\bf S}}{\partial {\bf U}}$ are $\left\{0,\,\lambda^+,\,\lambda^-\right\}$. If $m \leq n/2$, then $\Re e(\lambda^\pm) \leq 0$ and (\ref{SplitB}) is stable. This constraint is satisfied when the volume of the resonators is large compared to that of the necks, which is the case in practice (a similar argument has been used in section \ref{SecKdv});

\item[(ii)] linear attenuation ($m=n=0$), viscothermic losses in the waveguide ($c\neq0$) but not in the resonators ($f=0$). The eigenvalues of ${\bf T}$ are $\left\{0,\,\pm i\sqrt{g+eh},\,-\theta_\ell^2 \right\}$ hence (\ref{SplitB}) is stable;
\item[(iii)] linear attenuation ($m=n=0$), viscothermic losses ($c\neq 0$ and $f\neq 0$). Then 0 and $-\theta_\ell^2$ are simple eigenvalues of ${\bf T}$ ($\ell=1\cdots N$). Moreover, assuming positive weights $\mu_\ell>0$ and nodes $\theta_\ell>0$, and ordering the nodes as $0<\theta_1<\theta_2<\cdots<\theta_{N}$, then $N$ other eigenvalues $\lambda_\ell$ are real negative and satisfy:
\begin{equation}
\lambda_{N}<-\theta_{N}^2<\cdots<-\theta_{\ell+1}^2<\lambda_\ell<-\theta_{\ell}^2<\cdots<\lambda_1<-\theta_1^2<0.
\label{Encadre}
\end{equation}
In the limit-case $f=0$, the two remaining eigenvalues $\lambda_{2N+2}$ and $\lambda_{2N+3}$ are equal to the imaginary eigenvalues of case (ii): $\pm i\sqrt{g+eh}$. If $f\neq 0$, numerical tests indicate that these two eigenvalues are complex conjugate with a negative real part.
\end{itemize}
From cases (ii) and (iii), it follows that the spectral radius of the Jacobian satisfies $\varrho({\bf T})=|\lambda_N|>\theta_{N}^2\gg 1$. As a consequence, (\ref{SplitB}) must be solved by an implicit scheme \cite{Richoux15}. We conjecture that this stiffness of ${\bf T}$ still holds for nonlinear attenuation ($m\neq 0$ or $n \neq 0$) and for the general case (\ref{SystComplet}). This justifies the splitting strategy.

%---------------------------------------------------------------------------------

\section{Numerical results}\label{SecNum}

In this part, we examine whether the solution of the Sugiomoto's model (\ref{EDP}) has the typical features of solitons. In test 1, one investigates the dependence of the velocity upon the amplitude of the forcing. In test 2, we simulate the interaction between two waves.

\subsection{Study of the velocity in terms of the amplitude}

\begin{figure}[htbp]
\begin{center}
\begin{tabular}{cc}
(a) & (b)\\
\hspace{-0.8cm}
\includegraphics[width=8.8cm]{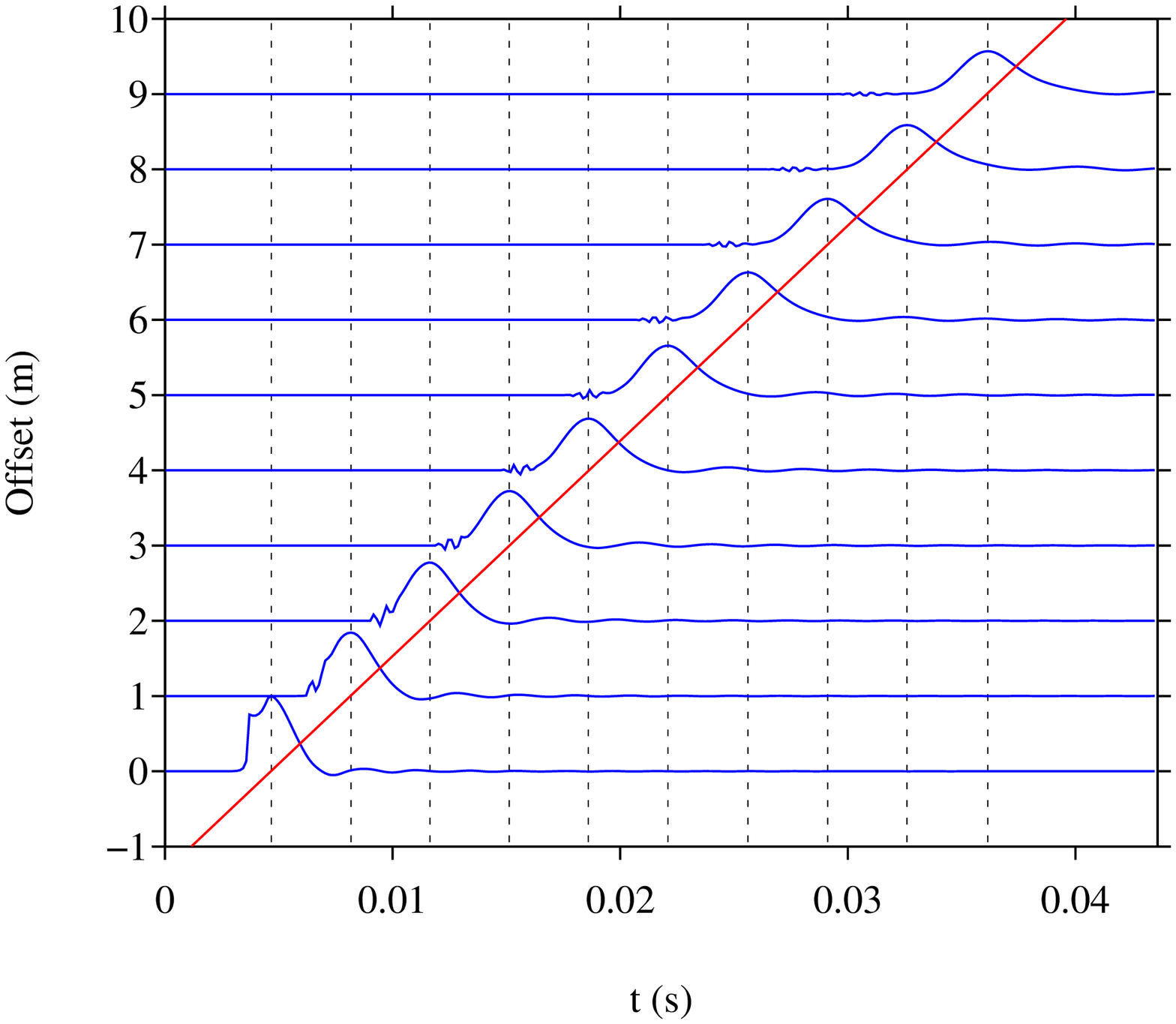} &
\hspace{-0.8cm}
\includegraphics[width=8.8cm]{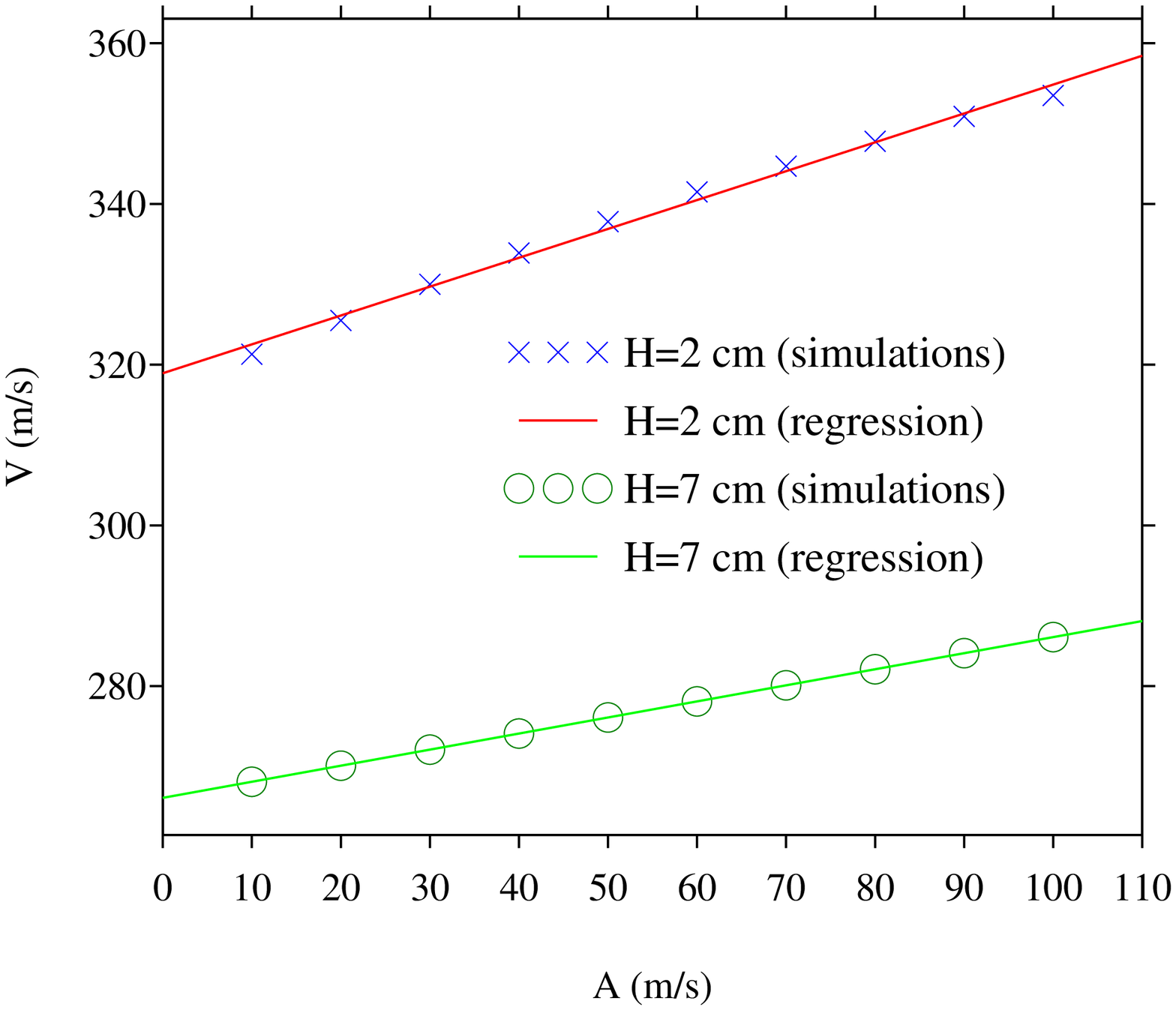} 
\end{tabular}
\end{center}
\caption{\label{FigSismo}Test 1. (a): example of seismogram. The vertical dotted lines represent the location of the maximum at each receiver. The inclined red line denotes the trajectory of these maxima; its slope yields the velocity of the wave. (b): velocity of the waves in terms of the forcing amplitude.} 
\end{figure}

The physical and geometrical parameters are given in \cite{Richoux15}. Two values of the resonators height are considered: $H=2$ cm and $H=7$ cm. This parameter influences the resonance angular frequency of the Helmholtz resonators ($\omega_e$ in section \ref{SecKdv}) and the parameters $K$ and $N$ in (\ref{Korteweg}). The waves are generated by imposing the value of the velocity in (\ref{SystComplet}) at $x=0$. A Gaussian with amplitude $A$ is chosen for this purpose:
\begin{equation}
u(0,t)=
\left\{
\begin{array}{l}
\ds
A\,e^{-\left(\frac{t-t_0}{\tau}\right)^2} \mbox{ if } 0\leq t \leq 2\,t_0,\\
[8pt]
\ds
0 \mbox{ otherwise}.
\end{array}
\right.
\label{Forcing}
\end{equation}
The central frequency is $f_0=1/t_0=650$ Hz. The standard deviation $\tau$ is chosen so that $u(0,0)=u(0,2\,t_0)=A/1000$. A set of 10 receivers is distributed uniformly on the computational domain. Seismograms are built from the time signals stored. The positions of the maximal value of $u$ at each receiver is detected and allows to estimate the celerity ${\cal V}$ of the wave. An example for $H=7$ cm and $A=100$ m/s is given in figure \ref{FigSismo}-(a). After a transient regime (offsets 0 and 1), a smooth structure emerges despite the nonsmoothness of the evolution equations (\ref{SystComplet}). The amplitude of the wave decreases along propagation, due to the loss mechanisms. Lastly, small amplitude waves are observed before the main wave front.

The same procedure is followed by varying $A$ from 10 m/s to 100 m/s. The evolution of ${\cal V}$ in terms of $A$ is illustrated in figure \ref{FigSismo}-(b). The linear increase of ${\cal V}$ with $A$ is clearly observed. Greater values of ${\cal V}$ are obtained for smaller value of $H$. These two observations confirm the theoretical analysis performed in \cite{Sugimoto04}. 

%---------------------------------------------------------------------------------

\subsection{Interaction of two solitons}

\begin{figure}[htbp]
\begin{center}
\begin{tabular}{cc}
(a) & (b)\\
\hspace{-0.8cm}
\includegraphics[width=8.8cm]{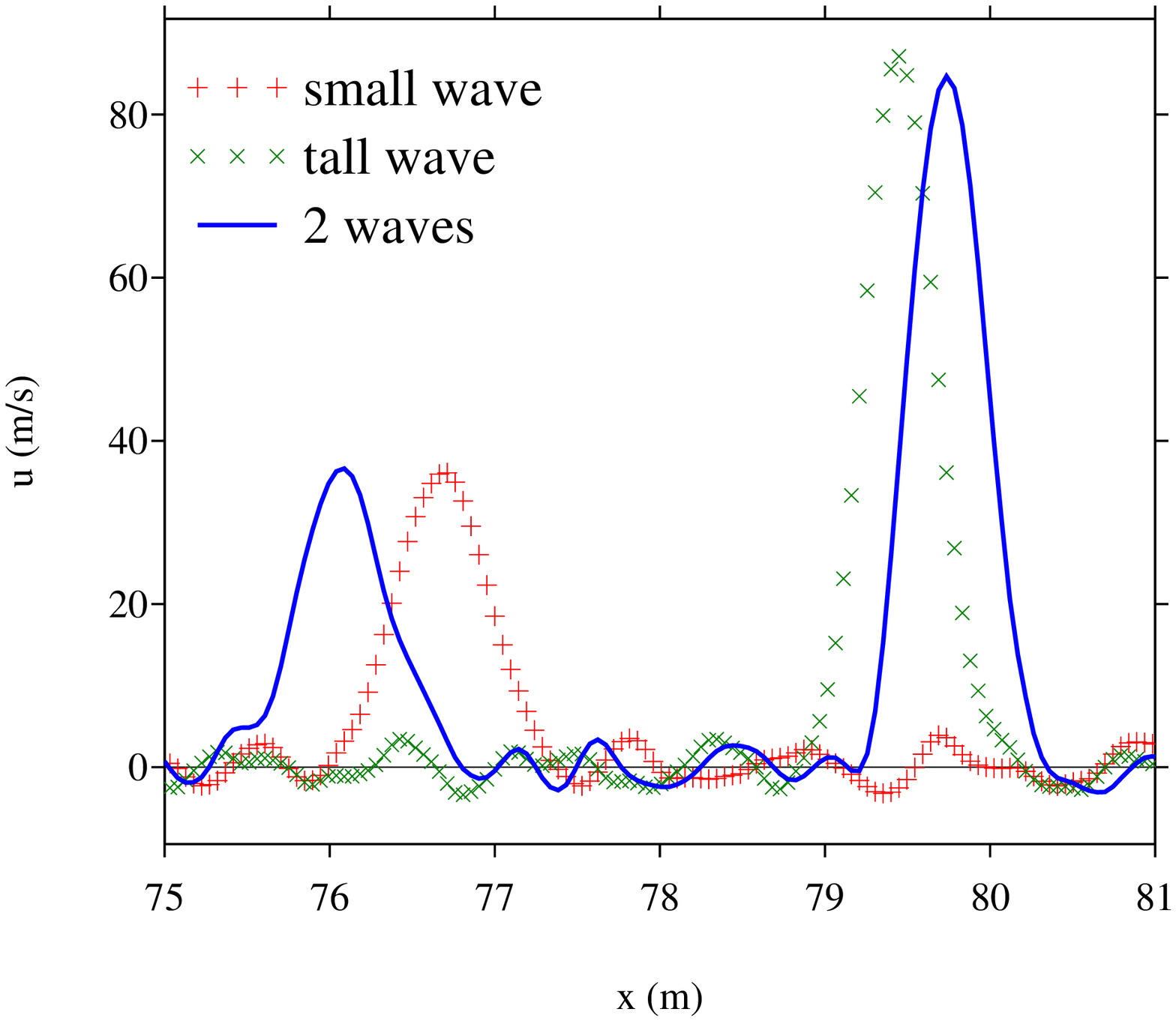} &
\hspace{-0.8cm}
\includegraphics[width=8.8cm]{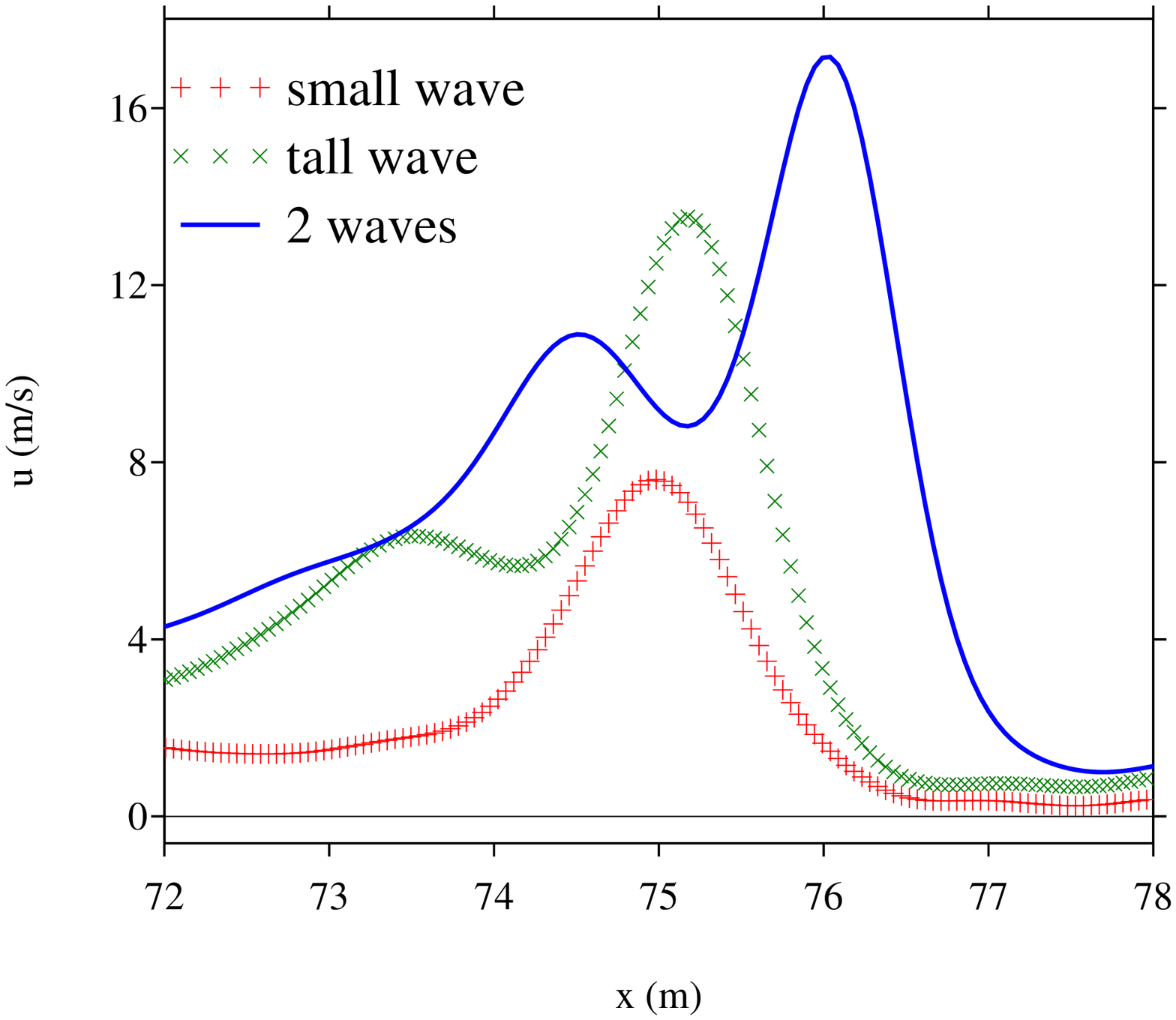} 
\end{tabular}
\end{center}
\caption{\label{FigLH}Test 2, with $H=7$ cm. Snapshots of $u$ after the interaction of two waves. (a): without dissipation. (b): with attenuation.} 
\end{figure}

A Gaussian pulse with small amplitude followed by a taller pulse are generated. Due to its higher amplitude, the latter travels faster, allowing an interaction between the two soliton waves. Figure \ref{FigLH} presents the results of this experiment. In the inviscid case (a), we observe that the two waves interact in a manner analogous to classical solitons \cite{LeVeque03}: after the waves separate, each one has again the form of a solitary wave, though shifted in location from where they would be without interaction (denoted by crosses). When the attenuation mechanisms are accounted for (b), a similar observation can be done, even if the observation is not so clear due to the smoothing of waves.

%---------------------------------------------------------------------------------
%---------------------------------------------------------------------------------

\vspace{-0.2cm}

\section{Conclusion}\label{SecConclu}

In this contribution, we have studied some properties of the full Sugimoto's model with nonlinear attenuation. Theoretical analysis has shown that the coefficient $m$ can produce problems (increase of energy, loss of stability). Since this coefficient has a negligible practical influence, we propose to remove it. Numerical experiments have allowed to examine situations difficult to reproduce experimentally. They have shown that typical features of solitons are maintained despite the nonlinear attenuation mechanisms.

\vspace{0.5cm}
\noindent
{\bf Acknowledgments}. This study has been supported by the Agence Nationale de la Recherche through the grant ANR ProCoMedia, project ANR-10-INTB-0914. The participation to IUTAM congress was funded by the French ANR 2010-G8EX-002-03 and by the foundation Del Duca, under grant 095164. 

%--------------------------------------------------------------------------

\end{document}